# Superconductivity in Potassium-intercalated $T_d$-WTe$_2$

Li Zhu[1], Qi-Yuan Li[1], Yang-Yang Lv[2], Shichao Li[1], Xin-Yang Zhu[1], Zhen-Yu Jia[1], Y. B. Chen[1,3]*, Jinsheng Wen[1,3]*, Shao-Chun Li[1,3]*

[1] National Laboratory of Solid State Microstructures, School of Physics, Nanjing University, Nanjing 210093, China

[2] National Laboratory of Solid State Microstructures, Department of Materials Science and Engineering, Nanjing University, Nanjing 210093, China

[3] Collaborative Innovation Center of Advanced Microstructures, Nanjing University, Nanjing 210093, China

Corresponding Author: scli@nju.edu.cn. jwen@nju.edu.cn. ybchen@nju.edu.cn.

**To realize topological superconductor is one of the most attracting topics because of its great potential in quantum computation. In this study, we successfully intercalate potassium ($K$) into the van der Waals gap of type II Weyl semimetal WTe$_2$, and discover the superconducting state in K$_x$WTe$_2$ through both electrical transport and scanning tunneling spectroscopy measurements. The superconductivity exhibits an evident anisotropic behavior. Moreover, we also uncover the coexistence of superconductivity and the positive magneto-resistance state. Structural analysis substantiates the negligible lattice expansion induced by the intercalation, therefore suggesting K-intercalated WTe$_2$ still hosts the topological nontrivial state. These results indicate that the K-intercalated WTe$_2$ may be a promising candidate to explore the topological superconductor.**

Various methods, such as high pressure [1], atomic intercalation [2,3], chemical substitution [4], and point contact [5], have been applied to tune the topological materials into superconducting state. $WTe_2$ has been extensively studied recently [6-14], particularly as the candidate of type-II Weyl semimetal [6-8] and due to the unsaturated magnetoresistance [9,10]. Interestingly, it has been found that bulk $WTe_2$ under high pressure exhibits the superconductivity with an optimal transition temperature ($T_c$) of ~6-7 K [15,16], however, there is no superconducting transition observed down to ~0.3 K at ambient pressure [17]. It is promising to explore the superconductivity in this system since the superconductivity and topological properties may coexist [18]. Several mechanisms have been proposed to explain the superconductivity transition in $WTe_2$ under high pressure, such as a quantum phase transition with the Fermi surface reconstruction due to compression [15], the enrichment of the density of state $N(E_F)$ in low-pressure regime and the possible structure instability at high pressure [16]. Later, it was attributed to the phase transition from ambient Td to the monoclinic 1T' phase [19]. However, in-depth understanding of the superconductivity in $WTe_2$ remains elusive, and the main challenge is the insufficient access of experimental techniques under high pressure. Therefore, to induce the superconductivity in $WTe_2$ at ambient pressure is significantly required.

Electron doping through alkali intercalation is an efficient way to tune superconductivity in transition-metal dichalcogenides (TMDs) [20-23], organic compounds [24,25], and iron-based superconductors [26-30]. Recently, ionic liquid gating has been widely applied to induce unconventional superconductivity in 2H phase TMD of $MX_2$ (M=W,

Mo; X=S, Se, Te) [31-35]. Here, we have for the first time successfully synthesized the K-intercalated $T_d$-WTe$_2$ by using liquid ammonia method [36, 37]. Resistance measurement demonstrated the superconductivity transition with Tc of ~ 2.6 K. By using scanning tunneling microscopy /spectroscopy (STM / STS), we found the local existence of SC gap at even higher temperature (~4.2 K) which may be attributed to the spatial inhomogeneity during the K intercalation. Unlike the superconductivity under high pressure [15], the positive magnetoresistance is still persisted in the superconducting K-intercalated WTe$_2$ samples, which may suggest an alternate mechanism. XRD measurement indicates that no prominent lattice expansion is detected, thus the emergence of superconductive transition can be mainly attributed to the electron doping effect.

Single crystal WTe$_2$ was grown by chemical vapor transport method. The K-intercalated WTe$_2$ sample was obtained through liquid ammonia method. More detailed description of liquid ammonia method is clarified in Supplementary materials. Single crystal WTe$_2$ was intercalated for the resistance measurement. Electrical resistance measurements were conducted in a physical property measurement system (PPMS) from Quantum Design. STM and STS measurements were carried out in a low temperature scanning tunneling microscope (LT-STM, Unisoku Co.) at ~4 K in ultra-high vacuum (UHV), with the base pressure of $1\times10^{-10}$ mbar. The K$_x$WTe$_2$ sample was *in situ* cleaved in UHV at room temperature, and then quickly transferred to STM stage for scan. For the powder X-ray diffraction characterization, the single crystal WTe$_2$ sample was grinded into powder prior to the intercalation process. Rietveld refinement was used to determine the crystal structure of $K_x$WTe$_2$.

Bulk WTe$_2$ possesses a layered structure of Td phase with the space group of $P_{nm21}$. W atomic layer is surrounded by up and down Te layers, forming sandwich structure. The Te-W-Te sandwich layer is stacked via van der Waals interactions, as schematically illustrated in Figure 1(a). The weak interlayer coupling allows for the entrance of alkali metal atoms or ions. The crystal quality of K-intercalated WTe$_2$ is characterized by XRD measurement. Figure 1(b) shows the powder XRD results measured on the sample with nominal $x$ = ~0.37. Rietveld refinement result from the Powder x-ray diffraction (PXRD) pattern (Fig. 1(b)) indicates that the $K_x$WTe$_2$ keeps the same crystal symmetry as the pristine WTe$_2$ [38], confirming that K is mostly intercalated inside the van de Waals gap. The Rietveld refined lattice parameters in the superconducting $K_{0.37}$WTe$_2$ are comparable with those obtained in the pristine WTe$_2$, as shown in Supplementary Materials Table S1. This is consistent with the previous studies on $Cu_x$Bi$_2$Se$_3$ and $Sr_x$Bi$_2$Se$_3$ [39, 40], where the atomic intercalation was not found to prominently enlarge the van de Waals gap (along the $c$ axis). In the study of K doped FeSe, the c axis is in fact increased through the formation of a new phase of $K_x$Fe$_2$Se$_2$ [37]. These structural analysis verify that the potassium intercalation between the WTe$_2$ layers has negligible effect on the crystal structure. In other words, negligible chemical pressure is induced by the K intercalation.

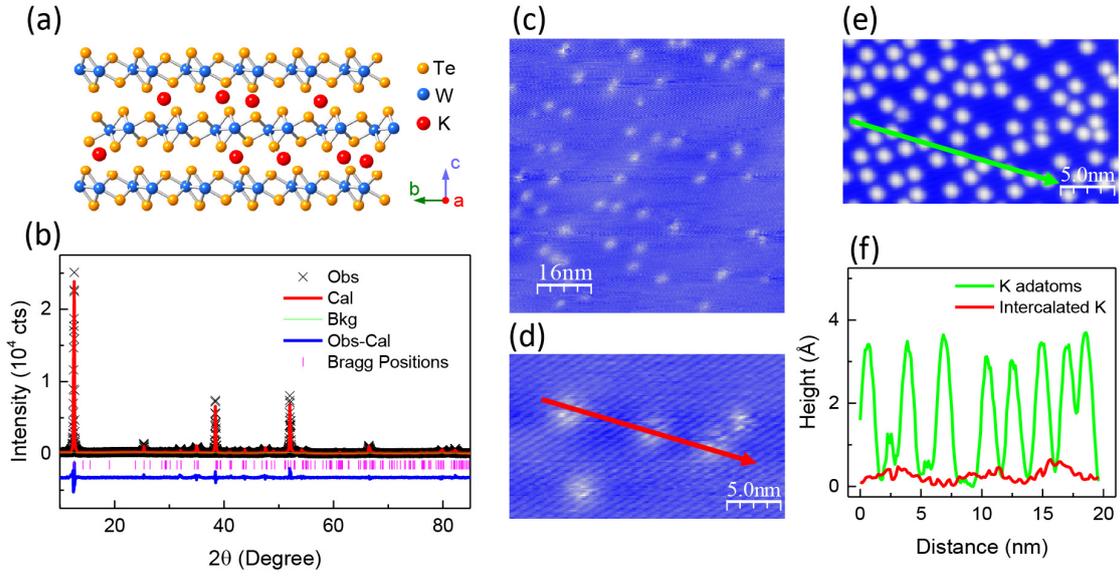

**Figure 1. Crystal structure and morphology of K-intercalated WTe$_2$.** (a) Schematic illustration of K being intercalated into the bulk WTe$_2$, K atoms are randomly distributed. (b) Powder XRD pattern (black cross mark) of the polycrystalline K$_{0.37}$WTe$_2$ and the well fitted Rietveld refinement results (red solid line). (c),(d) STM topographic images taken on the surface of K-intercalated WTe$_2$ (c: 80×80 nm$^2$, U = +200 mV, I$_t$=100 pA; d: 25×15 nm$^2$, U = +600 mV, I$_t$ = 80 pA ). (e) STM topographic image of the K-deposited WTe$_2$ surface (25×15 nm$^2$, U = +2 V, I$_t$ = 100 pA). (f) Line-cut profiles taken along the red and green arrowed lines in (d) and (e).

To verify the real-space morphology of the intercalated K in WTe$_2$, we performed the STM characterization on the cleaved surface of the K-intercalated sample. Figure 1(c) and (d) show the morphology of the K-intercalated WTe$_2$. Different from the pristine WTe$_2$ surface (in Supplementary Materials Fig. S1), there exist many small protrusions. In the highly resolved image shown in Figure 1(d), each protrusion exhibits with the patulous border, and resides over several lattice rows of WTe$_2$. Moreover, the surface lattice of WTe$_2$ can be well resolved on top of the protrusions. Line-cut profile taken across the protrusions, as plotted in Fig. 1(f), shows the quite small apparent height of ~ 50 pm. In comparison, we also deposited K atoms directly to the clean WTe$_2$ surface to achieve K adatoms. As shown in Figure 1(f), the apparent height of K adatom (green colored line in Fig. 1(e)) is much higher than that of the protrusions in K$_x$WTe$_2$ (red colored line in Fig. 1(d)), and the K adatom has well defined shape, different from the protrusions in K$_x$WTe$_2$. The possibility of surface Te vacancy can be also excluded due to the different STM morphology (see Supplementary Materials Fig. S2 for Te vacancy on clean WTe$_2$). Therefore, we assign these protrusions as the intercalated K atoms in the van der Waals gap underneath. Surface K adatoms are not observed on the surface of cleaved K$_x$WTe$_2$, which may be due to the fast surface diffusion and accumulation after the cleavage at room temperature. The WTe$_2$ surface was kept at liquid nitrogen temperature during K deposition to avoid the atom accumulation.

The electrical resistance (R) of K$_x$WTe$_2$ as a function of temperature (T) is plotted in Figure 2(a). The resistance shows a metallic behavior and tends to saturate below ~10 K, in agreement with the semi-metal nature of pristine WTe$_2$ [41, 42]. The resistance starts to take a sudden drop at ~2.6 K and then decreases to absolute zero at ~1.2 K, as can be clearly seen in the inset of Fig. 2(a). Here the transition temperature T$_c$ is defined as the onset point where the resistance starts to drop. To further confirm it is the signature of superconducting transition, the evolution of R-T curves under different magnetic field aligned along c-axis are presented in the inset of Figure 2(a). Obviously, the transition is gradually suppressed by the increasing magnetic field, and T$_c$ is shifted towards lower temperature. All of these results demonstrate that the K$_x$WTe$_2$ undergoes the superconducting transition at ~2.6 K.

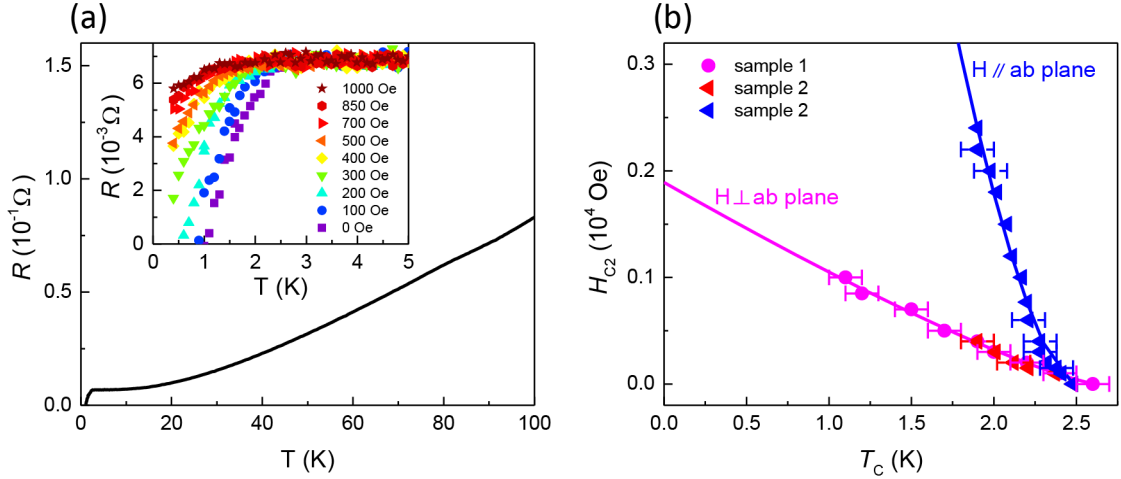

**Figure 2. Evidence of superconductivity in K$_x$WTe$_2$.** (a) Temperature dependence of the resistance (R) of K-intercalated WTe$_2$. Inset: The evolution of R vs T under different magnetic field H$_\perp$ perpendicular to the *ab*-plane. (b) The relation of transition temperature T$_c$ and magnetic field perpendicular (H$_\perp$) to or parallel (H$_\parallel$) with the *ab*-plane. The different symbols show the data taken on the different samples. The solid curves are the best fits by a simple power-law formula of $H_{c2}(T) = H_{c2}(0)(1 - T/T_C)^{1+\alpha}$.

The magneto resistance exhibits a strong anisotropic behavior when magnetic field is applied in different directions. In Fig. 2(b) are plotted the Tc vs magnetic field applied in both directions of perpendicular (H$_\perp$) to and parallel (H$_\parallel$) with the *ab* plane. The measured resistance results are shown in Supplementary Materials Fig. S3, for example, at T$_c$ = 2 K, the upper critical field H$_\parallel$ (~0.2 T) is nearly one order of magnitude larger than $H_\perp$ (~0.02 T). The experimental curve $H_{c2}(T_c)$ has a positive curvature near $T_{c0}$, which is consistent with the observation in the pressure-driven superconductivity of WTe$_2$ [16]. The temperature-dependent $H_{c2}$ can be well fitted with a simple empirical formula $H_{c2}(T) = H_{c2}(0)(1 - T/T_C)^{1+\alpha}$, and the upper critical field $H_{c2}(0)$ is estimated as ~1900 Oe for $H_\perp$ and ~23300 Oe for $H_{//}$, respectively. Noticeably, the obtained parameter α is 0.21 for $H_\perp$, comparable with the value for high pressurized WTe$_2$ [16] and 0.57 for $H_{//}$, comparable with MgB$_2$ [43] and LuPdBi [44]. According to the simple relationship between the coherent length and critical field, $H_{c2\perp} = \Phi_0/2\pi\xi_\parallel^2$, $H_{c2\parallel} = \Phi_0/2\pi\xi_\perp\xi_\parallel$, where $\Phi_0 = h/2e$, the coherence length $\xi_{//}$ can be estimated as ~41.6 nm, much larger than $\xi_\perp$ of ~3.4 nm, suggesting an anisotropic behavior in the superconducting K-intercalated WTe$_2$. Such anisotropic behavior of the upper critical field has been also revealed in other intercalated superconducting materials such as K$_x$Fe$_2$Se$_2$ [45] and Cu$_x$Bi$_2$Se$_3$ [2]. In the previous study, WTe$_2$ was also found to exhibit a similar anisotropy in the magneto-resistance study [9]. The origin resulting in such an anisotropic superconductivity may be ascribed to the highly anisotropic electronic structure of WTe$_2$.

A series of K$_x$WTe$_2$ samples with different nominal $x$ were explored. The curves of R vs T (Normalized to 200 K) are plotted together in Fig. 3(a). All the samples show a similar metallic behavior as the pristine WTe$_2$ at the non-superconducting regime. Superconducting transition is observed for the samples of $x$ = 0.32, 0.33 and 0.78. However, no such transition is observed for the samples of $x$ = 0 and 0.17 down to ~0.4

K. Therefore, the K doping level needs to be larger than a threshold value to realize superconductivity. In our experiment, the threshold value $x$ is between 0.17 and 0.32. Figure 3(b) shows the phase diagram of $T_C$ vs $x$. Once $x$ exceeds the threshold value, Tc increases quickly to the maximum value, and then decreases slowly with increasing $x$. The phase diagram is similar to that of $Cu_xBi_2Se_3$ [3], another possible candidate of topological superconductor.

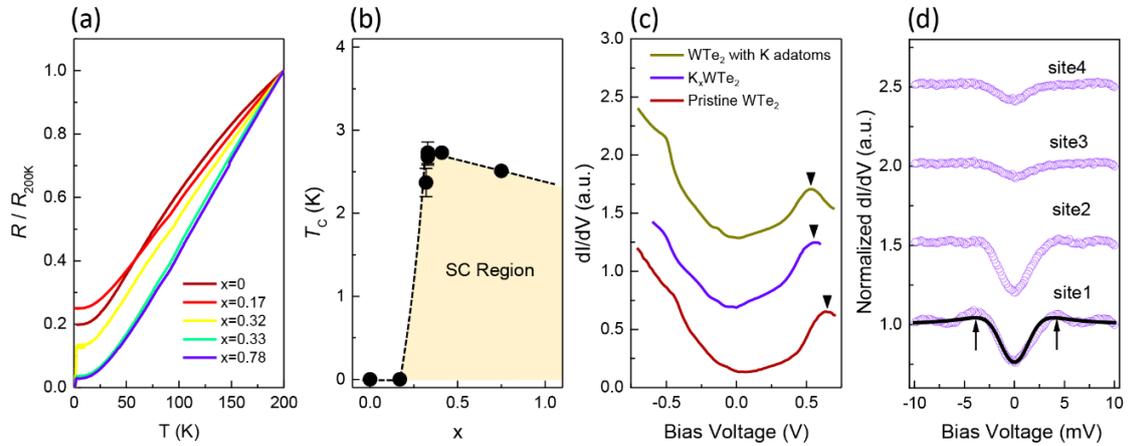

**Figure 3. The effect of potassium doping on the superconductivity in $K_xWTe_2$.** (a) Temperature dependence of the resistance (Normalized to 200 K) measured in a series of $K_xWTe_2$ with different nominal $x$. (b) Phase diagram of transition temperature $T_c$ vs doping level $x$. (c) dI/dV spectra (U = +600 mV, $I_t$ = 200 pA, $U_{mod}$ = 10 mV) measured in a large bias voltage scale (from -600 mV to +600 mV) on three different surface of pristine $WTe_2$, $K_xWTe_2$ and $WTe_2$ with K adatoms. (d) Normalized dI/dV spectra taken at different locations of the same surface of $K_xWTe_2$ (purple colored). The spectra are shifted by 0.5 in y axis for clarity. U = +10 mV, $I_t$ = 200 pA, $U_{mod}$ = 0.2 mV. The black solid curves show the BCS fitting.

To quantitatively characterize the electron doping level, dI/dV spectra (in bias range of ±600 mV) were taken on three different samples, the pristine WTe$_2$, K$_x$WTe$_2$ and WTe$_2$ with surface K adatoms for comparison. As plotted in Figure 3(c), the characteristic feature in the dI/dV curve of WTe$_2$, as marked by the black triangle, shifts with potassium doping level, implying that the Fermi energy shifts upwards due to doping electrons by K intercalation. For this specific sample of K$_x$WTe$_2$ ($x$ = 0.38), the Fermi energy is shifted by ~75 mV, in line with the surface doping of nominal 0.05 ML K. Thus, the main effect of K intercalation to WTe$_2$ is the electron doping, just like alkali metals do in the superconductivity of intercalated MoS$_2$[46].

Even though transport measurement shows an optimal Tc of ~2.6 K, STS measurement indicates the existence of local superconducting gap at even higher temperature. Figure 3(d) shows several typical dI/dV spectra taken at different areas of the same surface at ~4.2 K. Instead of an expected flat spectra, superconducting gaps with various size are observed in low bias. In contrast, the corresponding large scale (±600mV) spectra show consistent characteristic features along the surface. It is thus anticipated that the superconducting gap exhibits a drastic inhomogeneity: in some areas, the superconducting gap is small and almost smeared out (Site 3 and Site 4 of K$_x$WTe$_2$ in Fig. 3(d)); while in other areas, the superconducting gap is larger (Site 1 and Site 2 of K$_x$WTe$_2$ in Fig. 3(d)). Such surface inhomogeneity suggests that the local superconducting regions may still exist above Tc (~2.6 K) as determined in the transport result, but are not interconnected with each other. We ascribed such inhomogeneity of superconducting gap to the inhomogeneous doping effect, most likely the fluctuated local atomic concentration of K. In comparison, we also performed STS measurement on the NbSe$_2$ surface, a well characterized superconductor with Tc of ~7 K [47], with exactly the same instrumental set up (see supplementary materials Fig. S5). The gap of NbSe$_2$ [48, 49] is comparable with the maximum superconducting gap observed in the K$_x$WTe$_2$ sample. In principle, one can thus expect to realize a higher Tc of ~7 K in K$_x$WTe$_2$ by optimizing the electron doping effect.

The pronounced gap in K$_x$WTe$_2$ exhibits both characteristics of coherence peaks at the gap edge and the V shape inside. To explore the nature of the superconductivity in K$_x$WTe$_2$, we attempt to fit the STS results with BCS Dynes' model [50],

$$\frac{dI}{dV}(V) \cong Re\left[\frac{V-i\Gamma}{\sqrt{(V-i\Gamma)^2-\Delta^2}}\right], \qquad (1)$$

where $\Delta$ is the superconducting gap, $\Gamma$ is the effective energy broadening. A tentative best fitting to the superconducting gap (site1 in Fig. 3(d)) with a fully gapped s-wave function gives a $\Delta$ of ~1.6 mV and unreasonably large $\Gamma$ value of ~ 1.7 mV. Such a large $\Gamma$ value completely suppresses the coherence peak, and has no physical meaning. Even though to reduce the $\Gamma$ value can reproduce the coherence peak, but the calculated intensity at low bias is much lower than experiment. It is impossible to reproduce both features of the prominent coherence peak and the high in-gap intensity by tuning the values of $\Delta$ and $\Gamma$. Therefore, it is indicative of a not fully gaped superconducting states. A recent study of ionic liquid gated 2H MoS$_2$ [35] reported a resemble observation on the superconducting gap, which implies a mechanism beyond the conventional superconductivity. Furthermore, the superconducting inhomogeneity has been also observed in gated $WS_2$ [32]. Regardless of the different crystal structure, there may be a universal mechanism in this class of transition metal dichalcogenides $MX_2$ (M=W, Mo; X=S, Se, Te) materials.

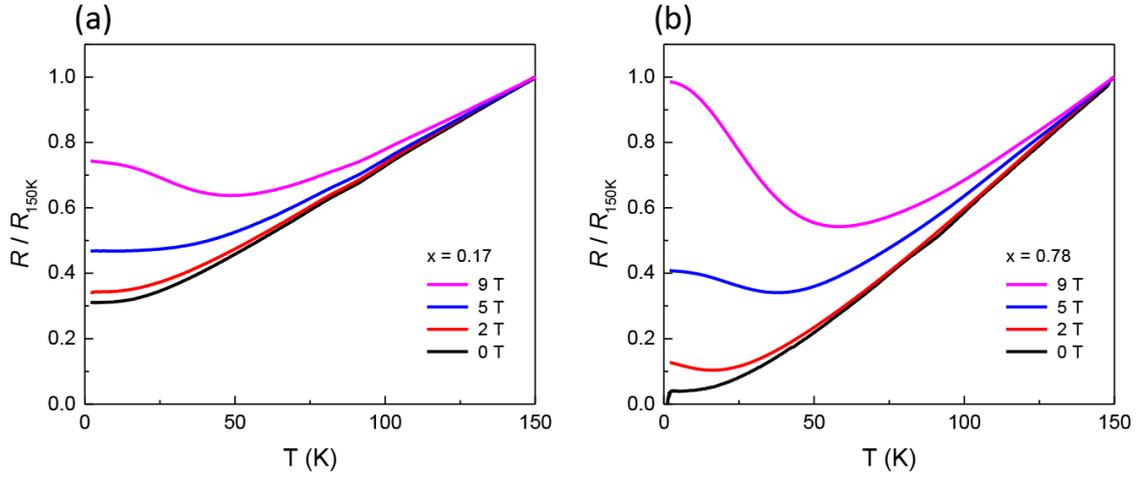

**Figure 4. The magneto-resistance of $K_xWTe_2$ samples.** (a), (b) The evolution of magneto resistance of non-superconducting $K_{0.17}WTe_2$ and superconducting $K_{0.78}WTe_2$ under a series of magnetic field perpendicular to the *ab* plane.

In general, atomic intercalation induces two effects of lattice expansion and electron doping. According to the XRD result (Fig. 1(b)) showing negligible lattice expansion and the STS results (Fig. 3(c)) showing similar characteristics, it is reasonable to assume a rigid band model for $K_xWTe_2$. K interaction donates electrons to the existing band and shift the Fermi energy. We thus believe the superconducting transition in $K_xWTe_2$ is mainly resulted from electron doping effect. In order to further compare with high-pressure superconductivity in $WTe_2$, another issue to investigate is the magneto-resistance effect in $K_xWTe_2$ under relatively high field in the non-superconducting regime. In the previous study [15], it was reported that the appearance of superconducting state is accompanied by the suppression of the large magnetoresistance state under high pressure. The temperature and field dependence of the magnetoresistance of two samples, non-superconducting $K_{0.17}WTe_2$ and superconducting $K_{0.78}WTe_2$, are measured and plotted in Fig. 4. Surprisingly, the positive magnetoresistance effect is always observable regardless of whether the sample is superconducting or not, in contrast with the previous conclusion made under high pressure [15]. In our study, it is clearly indicated that the positive magnetoresistance effect can coexist with the superconductivity in K-intercalated $WTe_2$. Under high pressure, the crystal lattice is compressed to result in the electronic structure change of $WTe_2$. While in the K intercalated $WTe_2$, superconductivity is promoted purely by electron doping effect, without a considerable lattice change. We therefore ascribe the suppression of MR to the band structure change which only applies in the high pressurized $WTe_2$. The superconductivity in $WTe_2$ can be induced purely by electron doping effect, and is not necessarily competitive with the large magneto resistance effect. Considering the structural intact in the superconducting $K_xWTe_2$, the non-trivial topological phase in $WTe_2$ may be still persisted.

In summary, superconductivity was successfully realized in $WTe_2$ at ambient pressure by K intercalated, with a Tc of ~2.6 K. Electron doping effect resulted from K-intercalation plays the dominant role in tuning the superconductivity. Since there is no change observed in the crystal lattice and no prominent suppression of the positive magnetoresistance, it is reasonable to expect that the topology of $WTe_2$ as the type II

Weyl semimetal is still persisted. K-intercalated WTe$_2$ thus provides a promising platform to explore the topological superconductor and the possible existence of Majorana Fermion.

ASSOCIATED CONTENT

**Supporting Information**.

The clarification of detailed experimental methods, comparison of *c*-direction lattice between pristine WTe$_2$ and intercalated WTe$_2$ (Table S1), XRD result of pristine single crystal WTe$_2$ (Figure S1), STM topographic image on the surface of pristine WTe$_2$ (Figure S2), resistance in K$_x$WTe$_2$ under magnetic field of H$_\perp$ and H$_\parallel$ (Figure S3). The color evolution of liquid ammonia (Figure S4). The STS gap of NbSe$_2$ (Figure S5).

**Acknowledgement:**

This work was supported by the Ministry of Science and Technology of China (Grants No. 2014CB921103, 2015CB921203), the National Natural Science Foundation of China (Grants No. 11774149, No. 11790311, 11674157, No. 11374149, No. 11374140), and the Open Research Fund Program of the State Key Laboratory of Low-Dimensional Quantum Physics.

# REFERENCES

(1) Kong, P. P.; Zhang, J. L.; Zhang, S. J.; Zhu, J.; Liu, Q. Q.; Yu, R. C.; Fang, Z.; Jin, C. Q.; Yang, W. G.; Yu, X. H.; Zhu, J. L.; Zhao, Y. S. *J. Phys.: Condens. Matter* **2013,** 25, (36), 5.

(2) Hor, Y. S.; Williams, A. J.; Checkelsky, J. G.; Roushan, P.; Seo, J.; Xu, Q.; Zandbergen, H. W.; Yazdani, A.; Ong, N. P.; Cava, R. J. *Phys. Rev. Lett.* **2010,** 104, (5), 057001.

(3) Kriener, M.; Segawa, K.; Ren, Z.; Sasaki, S.; Wada, S.; Kuwabata, S.; Ando, Y. *Phys. Rev. B* **2011,** 84, (5), 054513.

(4) Barzola-Quiquia, J.; Lauinger, C.; Zoraghi, M.; Stiller, M.; Sharma, S.; Haussler, P. *Supercond. Sci. Technol.* **2017,** 30, (1), 9.

(5) Wang, H.; Wang, H. C.; Liu, H. W.; Lu, H.; Yang, W. H.; Jia, S.; Liu, X. J.; Xie, X. C.; Wei, J.; Wang, J. *Nat. Mater.* **2016,** 15, (1), 38-43.

(6) Soluyanov, A. A.; Gresch, D.; Wang, Z.; Wu, Q.; Troyer, M.; Dai, X.; Bernevig, B. A. *Nature* **2015,** 527, (7579), 495-498.

(7) Zheng, H.; Bian, G.; Chang, G.; Lu, H.; Xu, S. Y.; Wang, G.; Chang, T. R.; Zhang, S.; Belopolski, I.; Alidoust, N.; Sanchez, D. S.; Song, F.; Jeng, H. T.; Yao, N.; Bansil, A.; Jia, S.; Lin, H.; Hasan, M. Z. *Phys. Rev. Lett.* **2016,** 117, (26), 266804.

(8) Lin, C. L.; Arafune, R.; Liu, R. Y.; Yoshimura, M.; Feng, B.; Kawahara, K.; Ni, Z.; Minamitani, E.; Watanabe, S.; Shi, Y.; Kawai, M.; Chiang, T. C.; Matsuda, I.; Takagi, N. *ACS Nano* **2017,** 11, (11), 11459-11465.

(9) Ali, M. N.; Xiong, J.; Flynn, S.; Tao, J.; Gibson, Q. D.; Schoop, L. M.; Liang, T.; Haldolaarachchige, N.; Hirschberger, M.; Ong, N. P.; Cava, R. J. *Nature* **2014,** 514, (7521), 205-208.

(10) Jiang, J.; Tang, F.; Pan, X. C.; Liu, H. M.; Niu, X. H.; Wang, Y. X.; Xu, D. F.; Yang, H. F.; Xie, B. P.; Song, F. Q.; Dudin, P.; Kim, T. K.; Hoesch, M.; Das, P. K.; Vobornik, I.; Wan, X. G.; Feng, D. L. *Phys. Rev. Lett.* **2015,** 115, (16).

(11) Liu, G.; Sun, H. Y.; Zhou, J.; Li, Q. F.; Wan, X.-G. *New J. Phys.* **2016,** 18, (3), 033017.

(12) Mleczko, M. J.; Xu, R. L.; Okabe, K.; Kuo, H. H.; Fisher, I. R.; Wong, H. S.; Nishi, Y.; Pop, E. *ACS Nano* **2016,** 10, (8), 7507-7514.

(13) Qian, X.; Liu, J.; Fu, L.; Li, J. *Science* **2014,** 346, (6215), 1344-1347.

(14) Jia, Z.-Y.; Song, Y.-H.; Li, X.-B.; Ran, K.; Lu, P.; Zheng, H.-J.; Zhu, X.-Y.; Shi, Z.-Q.; Sun, J.; Wen, J.; Xing, D.; Li, S.-C. *Phys. Rev. B* **2017,** 96, (4), 041108.

(15) Kang, D.; Zhou, Y.; Yi, W.; Yang, C.; Guo, J.; Shi, Y.; Zhang, S.; Wang, Z.; Zhang, C.; Jiang, S.; Li, A.; Yang, K.; Wu, Q.; Zhang, G.; Sun, L.; Zhao, Z. *Nat. Commun.* **2015,** 6, 7804.

(16) Pan, X. C.; Chen, X.; Liu, H.; Feng, Y.; Wei, Z.; Zhou, Y.; Chi, Z.; Pi, L.; Yen, F.; Song, F.; Wan, X.; Yang, Z.; Wang, B.; Wang, G.; Zhang, Y. *Nat. Commun.* **2015,** 6, 7805.

(17) Cai, P. L.; Hu, J.; He, L. P.; Pan, J.; Hong, X. C.; Zhang, Z.; Zhang, J.; Wei, J.; Mao, Z. Q.; Li, S. Y. *Phys. Rev. Lett.* **2015,** 115, (5), 057202.

(18) Fu, L.; Kane, C. L. *Phys. Rev. Lett.* **2008,** 100, (9), 4.

(19) Lu, P.; Kim, J.-S.; Yang, J.; Gao, H.; Wu, J.; Shao, D.; Li, B.; Zhou, D.; Sun, J.; Akinwande, D.; Xing, D.; Lin, J.-F. *Phys. Rev. B* **2016,** 94, (22), 224512.

(20) Somoano, R. B.; Rembaum, A. *Phys. Rev. Lett.* **1971,** 27, (7), 402-404.

(21) Onuki, Y.; Yamanaka, S.; Inada, R.; Kido, M.; Tanuma, S. *Synth. Met.* **1983,** 5, (3-4), 245-255.

(22) Morosan, E.; Wagner, K. E.; Zhao, L. L.; Hor, Y.; Williams, A. J.; Tao, J.; Zhu, Y.; Cava, R. J. *Phys. Rev. B* **2010,** 81, (9), 094524.

(23) Yu, Y. J.; Yang, F. Y.; Lu, X. F.; Yan, Y. J.; Cho, Y. H.; Ma, L. G.; Niu, X. H.; Kim, S.; Son, Y. W.; Feng, D. L.; Li, S. Y.; Cheong, S. W.; Chen, X. H.; Zhang, Y. B. *Nat. Nanotechnol.* **2015,** 10, (3), 270-276.

(24) Mitsuhashi, R.; Suzuki, Y.; Yamanari, Y.; Mitamura, H.; Kambe, T.; Ikeda, N.; Okamoto, H.; Fujiwara, A.; Yamaji, M.; Kawasaki, N.; Maniwa, Y.; Kubozono, Y. *Nature* **2010,** 464, (7285), 76-79.

(25) Wang, X. F.; Liu, R. H.; Gui, Z.; Xie, Y. L.; Yan, Y. J.; Ying, J. J.; Luo, X. G.; Chen, X. H. *Nat. Commun.* **2011,** 2, 507.

(26) Tapp, J. H.; Tang, Z.; Lv, B.; Sasmal, K.; Lorenz, B.; Chu, P. C. W.; Guloy, A. M. *Phys. Rev. B* **2008,** 78, (6), 060505.

(27) Guo, J.; Jin, S.; Wang, G.; Wang, S.; Zhu, K.; Zhou, T.; He, M.; Chen, X. *Phys. Rev. B* **2010,** 82, (18), 180520.

(28) Wang, A. F.; Ying, J. J.; Yan, Y. J.; Liu, R. H.; Luo, X. G.; Li, Z. Y.; Wang, X. F.; Zhang, M.; Ye, G. J.; Cheng, P.; Xiang, Z. J.; Chen, X. H. *Phys. Rev. B* **2011,** 83, (6), 060512.

(29) Krzton-Maziopa, A.; Shermadini, Z.; Pomjakushina, E.; Pomjakushin, V.; Bendele, M.; Amato, A.; Khasanov, R.; Luetkens, H.; Conder, K. *J. Phys.: Condens. Matter* **2011,** 23, (5), 052203.

(30) Scheidt, E. W.; Hathwar, V. R.; Schmitz, D.; Dunbar, A.; Scherer, W.; Mayr, F.; Tsurkan, V.; Deisenhofer, J.; Loidl, A. *Eur. Phys. J. B* **2012,** 85, (8).

(31) Ye, J. T.; Zhang, Y. J.; Akashi, R.; Bahramy, M. S.; Arita, R.; Iwasa, Y. *Science* **2012,** 338, (6111), 1193-1196.

(32) Jo, S.; Costanzo, D.; Berger, H.; Morpurgo, A. F. *Nano Lett.* **2015,** 15, (2), 1197-1202.

(33) Shi, W.; Ye, J. T.; Zhang, Y. J.; Suzuki, R.; Yoshida, M.; Miyazaki, J.; Inoue, N.; Saito, Y.; Iwasa, Y. *Sci. Rep.* **2015,** 5, 10.

(34) Costanzo, D.; Jo, S.; Berger, H.; Morpurgo, A. F. *Nat. Nanotechnol.* **2016,** 11, (4), 339-345.

(35) Costanzo, D.; Zhang, H. J.; Reddy, B. A.; Berger, H.; Morpurgo, A. F. *Nat. Nanotechnol.* **2018,** 13, (6), 483-488.

(36) Ying, T.-P.; Wang, G.; Jin, S.-F.; Shen, S.-J.; Zhang, H.; Zhou, T.-T.; Lai, X.-F.; Wang, W.-Y.; Chen, X.-L. *Chin. Phys. B* **2013,** 22, (8), 087412.

(37) Ying, T.; Chen, X.; Wang, G.; Jin, S.; Lai, X.; Zhou, T.; Zhang, H.; Shen, S.; Wang, W. *J. Am. Chem. Soc.* **2013,** 135, (8), 2951-2954.

(38) Lv, Y. Y.; Li, X.; Zhang, B. B.; Deng, W. Y.; Yao, S. H.; Chen, Y. B.; Zhou, J.; Zhang, S. T.; Lu, M. H.; Zhang, L.; Tian, M.; Sheng, L.; Chen, Y. F. *Phys. Rev. Lett.* **2017,** 118, (9), 096603.

(39) Hor, Y. S.; Checkelsky, J. G.; Qu, D.; Ong, N. P.; Cava, R. J. *J. Phys. Chem. Solids* **2011,** 72, (5), 572-576.

(40) Liu, Z.; Yao, X.; Shao, J.; Zuo, M.; Pi, L.; Tan, S.; Zhang, C.; Zhang, Y. *J. Am. Chem. Soc.* **2015,** 137, (33), 10512-10515.

(41) Lee, C. H.; Cruz-Silva, E.; Calderin, L.; Nguyen, M. A. T.; Hollander, M. J.; Bersch, B.; Mallouk, T. E.; Robinson, J. A. *Sci. Rep.* **2015,** 5, 8.

(42) Lu, N.; Zhang, C. X.; Lee, C. H.; Oviedo, J. P.; Nguyen, M. A. T.; Peng, X.; Wallace, R. M.; Mallouk,

T. E.; Robinson, J. A.; Wang, J. G.; Cho, K.; Kim, M. J. *J. Phys. Chem. C* **2016,** 120, (15), 8364-8369.
(43) Muller, K. H.; Fuchs, G.; Handstein, A.; Nenkov, K.; Narozhnyi, V. N.; Eckert, D. *J. Alloys Compd.* **2001,** 322, (1-2), L10-L13.
(44) Pavlosiuk, O.; Kaczorowski, D.; Wisniewski, P. *Sci. Rep.* **2015,** 5, 9.
(45) Ying, J. J.; Wang, X. F.; Luo, X. G.; Wang, A. F.; Zhang, M.; Yan, Y. J.; Xiang, Z. J.; Liu, R. H.; Cheng, P.; Ye, G. J.; Chen, X. H. *Phys. Rev. B* **2011,** 83, (21).
(46) Somoano, R. B.; Hadek, V.; Rembaum, A. *J. Chem. Phys.* **1973,** 58, (2), 697-701.
(47) Revolinsky, E.; Spiering, G. A.; Beerntsen, D. J. *J. Phys. Chem. Solids* **1965,** 26, (6), 1029-1034.
(48) Wang, M. X.; Liu, C. H.; Xu, J. P.; Yang, F.; Miao, L.; Yao, M. Y.; Gao, C. L.; Shen, C. Y.; Ma, X. C.; Chen, X.; Xu, Z. A.; Liu, Y.; Zhang, S. C.; Qian, D.; Jia, J. F.; Xue, Q. K. *Science* **2012,** 336, (6077), 52-55.
(49) Noat, Y.; Silva-Guillen, J. A.; Cren, T.; Cherkez, V.; Brun, C.; Pons, S.; Debontridder, F.; Roditchev, D.; Sacks, W.; Cario, L.; Ordejon, P.; Garcia, A.; Canadell, E. *Phys. Rev. B* **2015,** 92, (13), 134510.
(50) Dynes, R. C.; Narayanamurti, V.; Garno, J. P. *Phys. Rev. Lett.* **1978,** 41, (21), 1509-1512.